\documentclass[letterpaper,journal,twocolumn]{IEEEtran}
\usepackage{amsmath,amssymb}
\usepackage{mathrsfs}
\usepackage{amsthm}
\usepackage{graphicx,color,psfrag}
\usepackage{cite}

\usepackage{calc}
\graphicspath{{figs/}{matlab/}{../figs/}}

\newcommand{\Rbc}{\Rr_\mathrm{BC}}
\newcommand{\Rbcrelaxed}{\underline{\Rr}_\mathrm{BC}}
\newcommand{\Rmac}{\Rr_\mathrm{MAC}}
\newcommand{\Rmacdms}{\Rr'_\mathrm{MAC}}

\newcommand{\Rbcdms}{\Rr'_\mathrm{BC}}
\newcommand{\Ozgur}{\"Ozg\"ur}
\newcommand{\Rcs}{R_\mathrm{CS}}

\newcommand{\Rpdf}{R_\mathrm{PDF}}
\newcommand{\Rmpdf}{R'_\mathrm{PDF}}
\newcommand{\Rmpdfrelaxed}{R''_\mathrm{PDF}}
\newcommand{\Rmackdms}{\Rr'_\mathrm{MMAC}}

\newcommand{\Rmpdfsub}{R''_\mathrm{PDF}}
\newcommand{\Rddf}{R_\mathrm{DDF}}
\newcommand{\Deltaprimepdf}{\Delta{\hspace{-1pt}'}_\mathrm{\hspace{-1pt}PDF}}

\newtheorem{theorem}{Theorem}
\newtheorem{proposition}{Proposition}
\newtheorem{lemma}{Lemma}

\theoremstyle{remark}
\newtheorem{example}{Example}

\usepackage{xspace}
\usepackage{bbm}
\input{mathlig}

\newcommand{\muspace}{\mspace{1mu}}

\DeclareRobustCommand{\scond}{\mathchoice{\muspace\vert\muspace}{\vert}{\vert}{\vert}}
\mathlig{|}{\scond}

\DeclareRobustCommand{\discint}{\mathchoice{\mspace{-1.5mu}:\mspace{-1.5mu}}{\mspace{-1.5mu}:\mspace{-1.5mu}}{:}{:}}
\mathlig{::}{\discint}
\newcommand{\suchthat}{\mathchoice{\colon}{\colon}{:\mspace{1mu}}{:}}

\newcommand{\Jc}{\mathcal{J}}
\newcommand{\Kc}{\mathcal{K}}

\renewcommand{\Pr}{\mathscr{P}}
\newcommand{\Rr}{\mathscr{R}}

\newcommand{\pen}{{P_e^{(n)}}}

\newcommand{\Mh}{{\hat{M}}}

\newcommand{\Zh}{{\hat{Z}}}

\newcommand{\mh}{{\hat{m}}}

\newcommand{\Mt}{{\tilde{M}}}

\newcommand{\St}{{\tilde{S}}}

\newcommand{\Xt}{{\tilde{X}}}
\newcommand{\Yt}{{\tilde{Y}}}
\newcommand{\Zt}{{\tilde{Z}}}

\newcommand{\xt}{{\tilde{x}}}
\newcommand{\yt}{{\tilde{y}}}

\newcommand{\gt}{{\tilde{g}}}

\def\a{\alpha}
\def\b{\beta}

\DeclareMathOperator\E{\textsf{E}}
\let\P\relax
\DeclareMathOperator\P{\textsf{P}}

\DeclareMathOperator\C{\textsf{C}}

\def\error{\mathrm{e}}

\newcommand{\N}{\mathrm{N}}

\def\textiid{i.i.d.\@\xspace}
\newcommand\iid{\ifmmode\text{ i.i.d. } \else \textiid \fi}

\newcommand{\sfrac}[2]{\mbox{\small$\displaystyle\frac{#1}{#2}$}}
\newcommand{\half}{\sfrac{1}{2}}

\def\mathllap{\mathpalette\mathllapinternal}
\def\mathllapinternal#1#2{%
  \llap{$\mathsurround=0pt#1{#2}$}}

\def\clap#1{\hbox to 0pt{\hss#1\hss}}
\def\mathclap{\mathpalette\mathclapinternal}
\def\mathclapinternal#1#2{%
  \clap{$\mathsurround=0pt#1{#2}$}}

\let\oldstackrel\stackrel
\renewcommand{\stackrel}[2]{\oldstackrel{\mathclap{#1}}{#2}}

\renewcommand{\hbar}{h\mathllap{\overline{\vphantom{h}\hphantom{\rule{4.6pt}{0pt}}}\mspace{0.77mu}}}

\catcode`~=11 % make LaTeX treat tilde (~) like a normal character
\newcommand{\urltilde}{\kern -.06em\lower -.06em\hbox{~}\kern .02em}
\catcode`~=13 % revert back to treating tilde (~) as an active character

\title{Partial Decode--Forward Relaying\\ for
the Gaussian Two-Hop Relay Network}

\author{Jing Li and Young-Han Kim

\thanks{The material in this paper was presented in part at the 52nd Annual Allerton Conference on Communication, Control, and Computing, Monticello, Illinois, October 2014.

Jing Li is with the School of Telecommunications Engineering, Xidian University, Xi'an, Shaanxi 710071, China (email: jli\_8713@stu.xidian.edu.cn) and has been visiting the University of California, San Diego since September 2013. Young-Han Kim is with the Department of Electrical and Computer Engineering, University of California, San Diego, La Jolla, CA 92093,
USA (email: yhk@ucsd.edu).
}
}

\begin{document}

\maketitle

\begin{abstract}
The multicast capacity of the Gaussian two-hop relay network with one source, $N$ relays, and $L$ destinations is studied.
It is shown that a careful modification of the partial decode--forward coding scheme, whereby the relays cooperate
through degraded sets of message parts, achieves the cutset upper bound within $(1/2)\log N$ bits regardless of the channel gains and power
constraints. This scheme improves upon a previous scheme by Chern and \Ozgur, which is also based on partial decode--forward
yet has an unbounded gap from the cutset bound for $L \ge 2$ destinations.
When specialized to independent codes among relays,
the proposed scheme achieves within $\log N$ bits from the cutset bound. The computation of this relaxation involves evaluating mutual information across $L(N+1)$ cuts out of the total $L 2^N$ possible cuts, providing a very simple linear-complexity algorithm to approximate the single-source multicast capacity of the Gaussian two-hop relay network.
%Moreover, a more refined study of the performance of the distributed decode--forward is also proposed,
%and shown to achieve the cutset bound within $\log N +\frac{1}{2}$ bits regardless of the number of destinations.
\end{abstract}

\IEEEpeerreviewmaketitle

\section{Introduction}
\label{Introduction}
\begin{figure}[!t]
\begin{center}
\footnotesize
\psfrag{X}{$X$~}
\psfrag{Z1}{$Z_1$}
\psfrag{Z2}{$Z_2$}
\psfrag{ZN}{$Z_N$}
\psfrag{Y1}[lc]{\!\!$Y_1$}
\psfrag{Y2}[lc]{\!\!$Y_2$}
\psfrag{YN}[lc]{\!\!$Y_N$}
\psfrag{Xt1}[cr]{$\Xt_1$\!}
\psfrag{Xt2}[cr]{$\Xt_2$\!}
\psfrag{XtN}[cr]{$\Xt_N$\!}
\psfrag{Zt1}{$\Zt_1$}
\psfrag{ZtL}{$\Zt_L$}
\psfrag{Yt1}{$\Yt_1$}
\psfrag{YtL}{$\Yt_L$}
\includegraphics[width=3in]{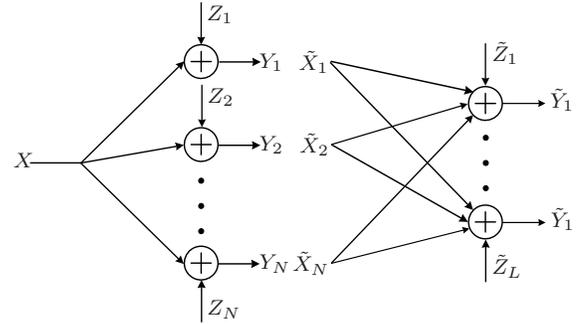}
\caption{The Gaussian two-hop relay network.}
\label{fig:multicast}
\end{center}
\end{figure}

Consider the Gaussian two-hop relay network with one source, $N$ relays, and $L$ destinations as depicted in Fig.~\ref{fig:multicast},
which can be viewed as a cascade of a broadcast channel (BC) from the source to the relays and multiple multiple access channels (MACs) from the relays to the destinations.
The source node wishes to reliably communicate a common message to the $L$ destination nodes
with help of the $N$ relays. The special case of $L = 1$, 
originally introduced by Schein and Gallager \cite{Schein--Gallager2000, Schein2001},
is often referred to as the \emph{diamond network}. The capacity is not known in general except for the trivial case of $N=1$.

The best known capacity upper bound is the cutset bound \cite{El-Gamal1981b}, which is the maximum of the minimum mutual information across all possible cuts that separate the source and the destinations.
There are several capacity lower bounds based on different coding schemes.
The compress--forward scheme for the 3-node relay channel by Cover and El Gamal \cite{Cover--El-Gamal1979} has been extended to relay networks in several forms,
such as quantize--map--forward (QMF) by Avestimehr, Diggavi, and Tse \cite{Avestimehr--Diggavi--Tse2011}, and noisy network coding (NNC) \cite{Lim--Kim--El-Gamal--Chung2011,Yassaee--Aref2011}. The standard analysis \cite{Lim--Kim--El-Gamal--Chung2011} shows that when specialized to our two-hop network model in Fig.~\ref{fig:multicast}, these coding schemes achieve the cutset bound within $O(N)$ bits for any channel parameters (recall that $N$ is the number of relays).

%The same performance can also be achieved by the standard analysis of distributed decode--forward (DDF) \cite{Lim--Kim--Kim--1--2014}.

Recently, Chern and \Ozgur \cite{Chern--Ozgur2012} provided a more refined analysis on the performance of NNC and showed that it achieves within $(1/2)\log 2N(N+1)$ bits from the cutset bound regardless of the number of destinations.
In the same paper \cite{Chern--Ozgur2012}, Chern and \Ozgur{} extended the partial decode--forward (PDF) scheme for the relay channel by Cover and El Gamal \cite{Cover--El-Gamal1979} to the Gaussian diamond network $(L=1)$. In the PDF scheme by Chern and \"Ozg\"ur, the source broadcasts independent parts of the message to the relays, which in turn recover and forward their corresponding parts to the destination over the MAC. Thus, the Chern--\"Ozg\"ur scheme can
achieve the rate
characterized by the intersection of the BC capacity region and the MAC capacity region, which can be shown to be
within $\log N$ bits
from the cutset bound. When there are more than one destination node, however, the gap from the cutset bound becomes unbounded \cite[Sec.~VI]{Chern--Ozgur2012}.

In this paper, we develop an alternative extension of partial decode--forward
that achieves the cutset bound within $\frac{1}{2}\log N$ bits for any number of destination nodes.
In the proposed scheme, the relays decode for multiple message parts based on their respective decoding capabilities
(as in the BC with \emph{degraded message sets} \cite{Bergmans1974})
and forward these parts cooperatively (as in the MAC with \emph{degraded message sets} \cite{Prelov1984, Han1979}).
Thus, the proposed scheme achieves the rate characterized by the intersection of the
capacity region of the BC with degraded message sets and the capacity regions
of the group of multiple access channels with degraded message sets.

Although this improvement may be viewed at first as an unnatural complication (except
for the obvious benefit for achieving higher multicast rates with $L \ge 2$ destinations), it actually
yields a simpler characterization of the achievable rate when independent Gaussian random codebooks are used at the relays, which yields
a slightly looser but easier-to-compute $\log N$ approximation of the capacity.
A direct computation of the cutset bound as well as of the achievable rates for NNC and the Chern--\"Ozg\"ur PDF scheme requires evaluating
mutual information across $L 2^N$ different cuts and then taking the minimum, which takes exponential
time when directly computed.
As an alternative to direct computation,
\emph{approximate} computation of the capacity (or the cutset bound) of the single-source single-destination
relay network has been proposed by Parvaresh and Etkin \cite{Parvaresh--Etkin2011} based on
properties of submodular function minimization,
which implies that the capacity of our two-hop network with $L = 1$
can be approximated within $2N$ in polynomial time of $O(LN^6)$ complexity (see also \cite{Brahma--Sengupta--Fragouli2014}).
In this paper,
we refine and strengthen the Parvaresh--Etkin approximation result by
showing that the achievable rate of our PDF scheme under independent codebooks involves evaluating only $L(N+1)$ cut rates. As a consequence,
we develop an explicit algorithm to approximate the capacity as well
as the cutset bound within $\log N$ with linear time complexity.

Finally, we evaluate the performance of yet another variant of partial decode--forward
for the two-hop relay network. Recently, Lim, Kim, and Kim developed distributed decode--forward, which generalizes partial decode--forward to general noisy networks for multicast \cite{Lim--Kim--Kim--1--2014} and broadcast \cite{Lim--Kim--Kim--2--2014}. As in the case of noisy network coding, a naive analysis of distributed decode--forward results in an achievable rate within $N/2$ bits from the cutset bound. In this paper,
we provide a refined analysis that establishes a gap of
$(\log N +\frac{1}{2})$ bits from the cutset bound.

The rest of the paper is organized as follows. In the next section,
we review basic facts on polymatroids.
In Section \ref{Model}, we formally define the capacity of the Gaussian two-hop relay network.
In Section~\ref{Cutset-Bound}, we review the cutset upper bound on the capacity, which will be benchmarked throughput.
In Section \ref{PDF}, we review the Chern--\Ozgur~partial decode--forward scheme for the Gaussian diamond network ($L = 1$).
In Section \ref{MPDF}, we present our coding scheme for the special case of the diamond network and then extend this result to the general $L$-destination case.
In Section \ref{Linear-complexity-app}, we show the computation of the achievable rate of the relaxed version of our coding scheme involves linear complexity.
%In Section \ref{Examples}, we provide some examples.
In Section \ref{DDF} and put forward the improved analysis of the performance of DDF.
Finally, we conclude the paper.

Throughout the paper, we mostly follow the notation in \cite{El-Gamal--Kim2011}.
In particular, we denote $[1:N]:=\{1,2,\cdots,N\}$.
The maximum of a finite set is denoted as $\Jc_{\max}:=\max(\Jc)$.
A tuple of random variables is denoted as $X(\Jc):=(X_j: j \in \Jc)$.
The Gaussian capacity function is defined as $\C(x) := (1/2) \log(1+x)$.

\section{Mathematical Preliminaries}
\label{Preliminaries}
Let $\phi: 2^{[1::N]}\to [0,\infty)$ be a set function satisfying
\begin{enumerate}
\item $\phi(\emptyset)=0$,
\item $\phi(\Jc)\le \phi(\Kc)$ if $\Jc \subseteq \Kc$, and
\item $\phi(\Jc \cap \Kc)+\phi(\Jc\cup \Kc) \le \phi(\Jc)+\phi(\Kc)$.
\end{enumerate}
Then the polyhedron
\begin{equation*}\begin{split}
\Pr(\phi) := \biggl \{ & (x_1,\cdots,x_N) \in [0,\infty)^N : \\
& \sum_{j \in \Jc} x_j \le \phi(\Jc),\, \Jc \subseteq [1::N] \biggr \}
\end{split}\end{equation*}
is said to be a \emph{polymatroid} (associated with $\phi$); see, for example,
\cite{Schrijver2003}.

\begin{example}
\label{polymatroid-1}
For any random tuple $(X_1, \ldots, X_N, Y)$ such that
$X_1, \ldots, X_N$ are mutually independent,
the set of rate tuples $(R_1, \ldots, R_N)$ satisfying
\[
\sum_{j \in \Jc} R_j \le I(X(\Jc); Y | X(\Jc^c))
\]
is a polymatroid \cite[Lemma~3.1]{Han1979}. In particular,
if $X_j \sim \N(0, S_j)$, $j \in [1::N]$, and $Y =
\sum_{j=1}^N X_j + Z$, where
$X_1, \ldots, X_N$
and $Z \sim \N(0,1)$
are mutually independent, then
the set of rate tuples $(R_1, \ldots, R_N)$ satisfying
\[
\sum_{j \in \Jc} R_j \le \textsf{\upshape{C}} \left( \sum_{j \in \Jc} S_j \right)
\]
is a polymatroid.
\end{example}

\begin{example}
\label{polymatroid-2}
Let $\Phi: [1:N] \to [0,\infty)$ be nondecreasing and define
$\phi: 2^{[1::N]}\to [0,\infty)$ by
\[
\phi(\Jc) =
\begin{cases}
0, & \Jc = \emptyset, \\
\Phi(\Jc_{\max}), & \text{otherwise}.
\end{cases}
\]
Then it can be readily shown that $\Pr(\phi)$ is a polymatroid characterized by active inequalities
\[
\sum_{j=1}^k x_j \le \phi([1:k]) = \Phi(k), \quad k \in [1::N].
\]
In particular,
for any random tuple $(X_1, \ldots, X_N, Y)$,
the set of rate tuples $(R_1, \ldots, R_N)$ satisfying
\[
\sum_{j=1}^k R_j \le I(X^k ; Y | X_{k+1}^N)
\]
is a polymatroid.
\end{example}

The following well-known result is pivotal in our discussion.

\begin{lemma}[Edmonds's polymatroid intersection theorem~\cite{Edmonds1970}]
\label{Edmond}
If $\Pr(\phi)$ and $\Pr(\psi)$ are two polymatroids, then
\begin{align*}
&\max \biggl \{\sum_{j=1}^N x_j : (x_1, \cdots, x_N) \in \Pr(\phi) \cup \Pr(\psi)  \biggr \} \\
&\qquad=\min_{\Jc \subseteq [1::N]} \bigl [ \phi(\Jc)+\psi(\Jc^c) \bigr ].
%&\max_{\Jc \subseteq [1::N]}\biggl \{\sum_{j\in \Jc} x_j : (x_1, \cdots, x_N) \in \Pr(\phi) \cup \Pr(\psi)  \biggr \} \\
%&\qquad=\min_{\Jc \subseteq [1::N]} \bigl [ \phi(\Jc)+\psi(\Jc^c) \bigr ].
\end{align*}
\end{lemma}

%
%\begin{corollary} If $\Pr(\phi)$ satisfies the conditions in Lemma \ref{polymatroid-1}, and $\Pr(\psi)$ is a polymatroid, then
%  \begin{align*}
%        &\max \biggl \{\sum_{j\in \Jc} x_j : (x_1, \cdots, x_N) \in \Pr(\phi) \cup \Pr(\psi)  \biggr \} \\
%        &=\min_{j \in [0::N]} \bigl [ \phi([1::j])+\psi([j+1::N]) \bigr ].
%  \end{align*}
%where $\phi([1::0])=\psi([N+1::N])=0$.
%\label{Edmond-1}
%\end{corollary}
%\emph{Proof}. According to Lemma \ref{Edmond},
%\begin{align*}
%&\max \biggl \{\sum_{j\in \Jc} x_j : (x_1, \cdots, x_N) \in \Pr(\phi) \cup \Pr(\psi)  \biggr \} \\
%&=\min_{\Jc \subseteq [1::N]} \bigl [ \phi(\Jc)+\psi(\Jc^c) \bigr ] \\
%&=\min_{\Jc \subseteq [1::N]} \bigl [ \phi([1::\Jc_{max}])+\psi(\Jc^c) \bigr ] \\
%&\le \min_{\Jc \subseteq [1::N]} \bigl [ \phi([1::\Jc_{max}])+\psi([\Jc_{max}+1::N]) \bigr ] \\
%&=\min_{j \in [1::N]} \bigl [ \phi([1::j])+\psi([j+1::N]) \bigr ]. \\
%\end{align*}

\section{Formal Definition of Capacity}
\label{Model}
Recall the Gaussian two-hop relay network model depicted in Fig.~\ref{fig:multicast}. 
The received signals at the relays corresponding to the signal $X$ from the source node are
\[
Y_j = g_j X + Z_j,\quad j \in [1:N],
\]
where $g_1, \ldots, g_N$ are the channel gains from the source to relay nodes~$1$ through $N$, respectively,
and $Z_1, \ldots, Z_N$ are independent $\N(0,1)$ noise components. We assume without loss of generality that
\begin{equation} \label{eq:gains}
|g_1| \ge |g_2| \ge \cdots \ge |g_N|.
\end{equation}
Similarly, the received signals at the destinations corresponding to the signals $\Xt_1, \ldots, \Xt_N$ transmitted from the relays are
\[
\Yt_{d} = \sum_{j=1}^N \gt_{dj} \Xt_j + \Zt_{d},\quad d \in [1::L],
\]
where $\gt_{dj}$, $j \in [1::N]$, $d \in [1::L]$, denote the channel gain from relay node $j$ to destination node $d$,
and $\Zt_{1}, \ldots, \Zt_{L}$ are independent $\N(0,1)$ noise components.
The first (source-to-relays) hop of the network can be viewed as a Gaussian broadcast channel, while the second (relays-to-destinations) hop of the network can be viewed as multiple Gaussian multiple access channels.
All nodes are subject to (expected) average power constraint $P$, and
we denote by $S_j = g_j^2 P$ and $\St_{dj} = \gt_{dj}^2 P$ the received
signal-to-noise ratios (SNRs) at the relays and the receivers,
respectively.

We define a $(2^{nR},n)$ code for a Gaussian two-hop relay network by
\begin{itemize}
\item a message set $[1::2^{nR}]$,

\item an source encoder that assigns a codeword $x^N(m)$ to each message $m\in [1::2^{nR}]$,

\item a set of relay encoders, where encoder $j\in [1::N]$ assigns a symbol $\xt_{ji}(y_{j}^{i-1})$ to each past received sequence $y_{j}^{i-1}$ for each transmission time $i \in [1::n]$, and

\item a set of decoders, where decoder $d \in [1::L]$ assigns an estimate $\mh_{d}$ or an error message $\error$ to each received sequence $\yt_{d}^N$.
\end{itemize}
We assume that the message $M$ is uniformly distributed over the message set. The average probability of error is defined as $\pen=\P\{\Mh_d\ne M \text{ for~some } d\in [1::L]\}$. A rate $R$ is said to be achievable for the Gaussian two-hop relay network if there exists a sequence of $(2^{nR},n)$ codes such that $\lim_{n \to \infty} \pen =0$. The capacity $C$ is defined as the supremum of all achievable rates.

When $N = 1$, the capacity is
\[
C = \min \biggl \{ \C(S_1), \, \min_d \C(\St_{d1}) \biggr \}.
\]
For $N \ge 2$, however, no computable characterization of the capacity is known even when $L = 1$.
In subsequent sections, we present bounds on the capacity and establish their closeness.

\section{The Cutset Bound on the Capacity}
\label{Cutset-Bound}
Since the network consists of two noninteracting channel layers,
the cutset bound~\cite{El-Gamal1981b} on the capacity of a general noisy network
can be simplified as
\begin{align}
C &\le \Rcs  \notag\\
&:=\sup_F \min_{d,\Jc}
\bigl [I(X; Y(\Jc^c)) + I(\Xt(\Jc); \Yt_{d} | \Xt(\Jc^c) ) \bigr],
\label{eq:cutset}
\end{align}
where the supremum is over
all joint distributions $F(x)F(\xt^N)$ satisfying $\E (X^2)\le P$ and $\E (\Xt_j^2)\le P$, $j\in [1::N]$,
the minimum is over all $d\in[1::L]$ and $\Jc \subseteq [1::N]$, and $\Jc^c$ denotes $[1::N] \setminus \Jc$.
By the maximum differential entropy lemma (see, for example, \cite[Section~2.2]{El-Gamal--Kim2011}),
the supremum in \eqref{eq:cutset} is attained by Gaussian $X$ and jointly Gaussian
$(\Xt_1,\ldots, \Xt_N)$. By switching the order of the supremum (over Gaussian distributions)
and the minimum, the cutset bound is further upper bounded as
\begin{align}
\Rcs &\le
\sup_{F(\xt^N)} \min_{d,\Jc} \sup_{F(x)}
\bigl [I(X; Y(\Jc^c)) + I(\Xt(\Jc); \Yt_{d} | \Xt(\Jc^c) ) \bigr ] \notag\\
&= \sup_{F(\xt^N)}\min_{d, \Jc}
\biggl [ \C\biggl( \sum_{j \in \Jc^c} S_j \biggr)
+ I(\Xt(\Jc); \Yt_{d} | \Xt(\Jc^c) ) \biggr ] \label{eq:cutset1} \\
&\le \min_{d, \Jc}\sup_{F(\xt^N)}
\biggl [ \C\biggl( \sum_{j \in \Jc^c} S_j \biggr)
+ I(\Xt(\Jc); \Yt_{d} | \Xt(\Jc^c) ) \biggr ] \notag \\
&\le \min_{d, \Jc}
\biggl [ \C\biggl( \sum_{j \in \Jc^c} S_j \biggr)
+ \C\biggl( \Bigl(\sum_{j\in \Jc} \sqrt{\St_{dj}}\Bigr)^2 \biggr) \biggr ]. \label{eq:cutset2}
\end{align}
Note that direct computation of the cutset bound in \eqref{eq:cutset1} for a fixed distribution or
its relaxation in \eqref{eq:cutset2}
involves evaluation of the minimum rate over the combination of $2^N$ choices of $\Jc$
and $L$ choices of $d$, that is, the total $L 2^N$ cuts that separate the source and the destinations.

\section{The Chern--\Ozgur{} Partial Decode--Forward Scheme for the Gaussian Diamond Network}
\label{PDF}
\begin{figure}[b]
\centering
\psfrag{M}{\tiny $M_1,M_2,...,M_N$}
\psfrag{A}{\tiny $X^{n}$}
\psfrag{A1}{\tiny $\Xt_{1}^{n}$}
\psfrag{A2}{\tiny $\Xt_{2}^{n}$}
\psfrag{AN}{\tiny $\Xt_{N}^{n}$}
\psfrag{B1}{\tiny $Y_{1}^{n}$}
\psfrag{B2}{\tiny $Y_{2}^{n}$}
\psfrag{BN}{\tiny $Y_{N}^{n}$}
\psfrag{Bt1}{\tiny $\Yt_{1}^{n}$}
\psfrag{C1}{\tiny $\Mt_1$}
\psfrag{C2}{\tiny $\Mt_2$}
\psfrag{CN}{\tiny $\Mt_N$}
\psfrag{D}{\tiny $\Mh_1,\Mh_2,...,\Mh_N$}
\psfrag{Enc}{\tiny Enc}
\psfrag{Dec}{\tiny Dec 1}
\psfrag{RE1}{\tiny Relay Enc 1}
\psfrag{RE2}{\tiny Relay Enc 2}
\psfrag{REN}{\tiny Relay Enc N}
\psfrag{RD1}{\tiny Relay Dec 1}
\psfrag{RD2}{\tiny Relay Dec 2}
\psfrag{RDN}{\tiny Relay Dec N}
\psfrag{BC}{\tiny BC}
\psfrag{MAC}{\tiny MAC}
\includegraphics[width=3.45in]{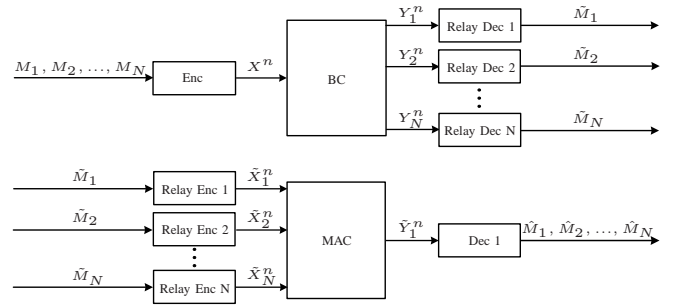}
\caption{The Chern--\Ozgur{} partial decode--forward coding scheme for $L = 1$.}
\label{fig:chern-ozgur}
\end{figure}

In the partial decode--forward scheme by Chern and \Ozgur~\cite{Chern--Ozgur2012} (see also \cite{Chern--Ozgur2014}), which was developed mainly for the case $N = 1$,
the source node divides the message $M$ into $N$ independent parts $M_1, \ldots, M_N$ (rate splitting), relay $j$ recovers $M_j$ and forwards it (decode--forward), and the destination node forms the estimates of $M_1,\ldots, M_N$ and thus of $M$ itself; see Fig.~\ref{fig:chern-ozgur}.
This scheme is implemented over two hops in a block Markov fashion, and the achievable rate can be characterized as
\begin{equation}
\label{eq:Rpdf}
\Rpdf =
\max \biggl\{ \sum_{j=1}^N R_j \suchthat (R_1, \ldots, R_N) \in \Rbc \cap \Rmac \biggr\}.
\end{equation}
Here $\Rbc$ is the capacity region of the standard $N$-receiver Gaussian broadcast channel with SNRs $S_1,\ldots, S_N$, that is,
the set of rate tuples $(R_1,\ldots, R_N)$ such that
\begin{equation} \label{eq:Rbc}
R_j\le \C \biggl(\frac{\alpha_j S_j}{\sum_{k=1}^{j-1}\alpha_k S_k+1} \biggr),\quad j \in [1::N],
\end{equation}
for some $(\a_1,\ldots, \a_N)$ satisfying $\a_j \ge 0$, $j \in [1::N]$, and $\sum_{j=1}^N \a_j = 1$,
which, by the BC--MAC duality \cite{Vishwanath--Jindal--Goldsmith2003},
can be written as the set of rate pairs $(R_1, \ldots, R_N)$ such that
\begin{equation}
\label{eq:dual-bc}
\sum_{j \in \Jc} R_j \le \C\biggl( \sum_{j \in \Jc} \b_j S_j \biggr),\quad
\Jc \subseteq [1::N],
\end{equation}
for some $(\b_1,\ldots, \b_N)$ satisfying $\b_j \ge 0$, $j \in [1::N]$,
and $\sum_{j=1}^N \b_j = 1$.
In~\eqref{eq:Rpdf}, $\Rmac$ is the capacity region of the standard $N$-sender Gaussian multiple access channel with SNRs $\St_{11}, \ldots, \St_{1N}$, i.e.,
the set of rate tuples $(R_1,\ldots, R_N)$ such that
\begin{equation*}
\sum_{j\in \Jc} R_j \le \C\biggl( \sum_{j\in \Jc} \St_{1j} \biggr),\quad
\Jc \subseteq [1::N].
\end{equation*}
Note that the region $\Rmac$ is a polymatroid (cf.\@ Example~\ref{polymatroid-1}), but the region $\Rbc$ is not in general. Consequently, the maximum sum-rate of the intersection of the two regions, characterized by \eqref{eq:Rpdf}, is rather cumbersome to calculate. Chern and \Ozgur{}
set $\b_j \equiv 1/N$ in~\eqref{eq:dual-bc} to obtain a \emph{polymatroidal} inner bound on $\Rbc$ characterized by
\begin{equation} \label{eq:bc-ineqs}
\sum_{j\in \Jc} R_j \le \C\biggl( \sfrac{1}{N} \sum_{j \in \Jc} S_j \biggr),\quad
\Jc \subseteq [1::N].
\end{equation}
Now by~\eqref{eq:Rpdf}
and Edmonds's polymatroid intersection theorem with
\begin{align*}
\phi(\Jc) &= \C\biggl( \sum_{j\in \Jc} \St_{1j} \biggr), \\
\psi(\Jc) &= \C\biggl( \sfrac{1}{N} \sum_{j \in \Jc} S_j \biggr),
\end{align*}
the corresponding (lower bound on the) achievable rate is
\begin{align}
\Rpdf &\ge
\min_{\Jc \subseteq [1::N]} \bigl [ \phi(\Jc) + \psi(\Jc^c) \bigr ] \notag \\
&= \min_{\Jc \subseteq [1::N]}
\biggl [
\C\biggl( \sfrac{1}{N} \sum_{j \in \Jc^c} S_j \biggr)
+ \C\biggl( \sum_{j\in \Jc} \St_{1j} \biggr)
\biggr ].
\label{eq:pdf}
\end{align}
By comparing this rate with the capacity upper bound in \eqref{eq:cutset2}, we
observe that the gaps for the two terms,
both due to the lack of coherent cooperation,
are bounded uniformly as
\begin{align}
\C\biggl( \sum_{j \in \Jc^c} S_j \biggr) - \C\biggl( \sfrac{1}{N} \sum_{j \in \Jc^c} S_j \biggr)
&\le \half \log N, \label{eq:gap-term1}\\
\C\biggl( \Bigl(\sum_{j\in \Jc} \sqrt{\St_{1j}}\Bigr)^2 \biggr)
- \C\biggl( \sum_{j\in \Jc} \St_{1j} \biggr) &\le \half \log N. \label{eq:gap-term2}
\end{align}
In conclusion,
the gap between the achievable rate of the Chern--\Ozgur{} partial decode--forward
scheme and the cutset bound is upper bounded as
\[
\Delta_\mathrm{PDF} := \Rcs - \Rpdf \le \log N,
\]
regardless of $S_j$ and $\St_{1k}$, $j, k \in [1::N]$.
%Note that computation of \eqref{eq:pdf} involves minimization
%over $2^N$ values. Consequently, the capacity
%is computed approximately within $\log N$ bits in polynomial time
%by considering $2^N$ cuts as in \eqref{eq:cutset2}.

\section{The Proposed Partial Decode--Forward Scheme}
\label{MPDF}
\begin{figure}[h]
\centering
\psfrag{M}{\tiny $M_1,M_2,...,M_N$}
\psfrag{A}{\tiny $X^{n}$}
\psfrag{A1}{\tiny $\Xt_{1}^{n}$}
\psfrag{A2}{\tiny $\Xt_{2}^{n}$}
\psfrag{AN}{\tiny $\Xt_{N}^{n}$}
\psfrag{B1}{\tiny $Y_{1}^{n}$}
\psfrag{B2}{\tiny $Y_{2}^{n}$}
\psfrag{BN}{\tiny $Y_{N}^{n}$}
\psfrag{Bt1}{\tiny $\Yt_{1}^{n}$}
\psfrag{C1}{\tiny $\Mt_1,\Mt_2,...,\Mt_N$}
\psfrag{C2}{\tiny $\Mt_2,...,\Mt_N$}
\psfrag{CN}{\tiny $\Mt_N$}
\psfrag{D}{\tiny $\Mh_1,\Mh_2,...,\Mh_N$}
\psfrag{Enc}{\tiny Enc}
\psfrag{Dec}{\tiny Dec 1}
\psfrag{RE1}{\tiny Relay Enc 1}
\psfrag{RE2}{\tiny Relay Enc 2}
\psfrag{REN}{\tiny Relay Enc N}
\psfrag{RD1}{\tiny Relay Dec 1}
\psfrag{RD2}{\tiny Relay Dec 2}
\psfrag{RDN}{\tiny Relay Dec N}
\psfrag{BC1}{\tiny BC with}
\psfrag{BC2}{\tiny degraded}
\psfrag{BC3}{\tiny message sets}
\psfrag{MAC1}{\tiny MAC with}
\includegraphics[width=3.45in]{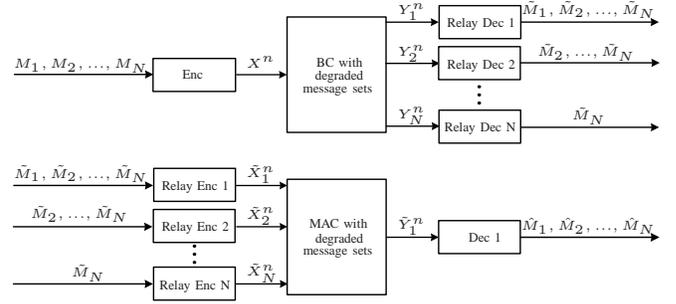}
\caption{The proposed partial decode--forward scheme for $L = 1$.}
\label{fig:scheme}
\end{figure}

We propose a modified version of the Chern--\Ozgur{} partial decode--forward scheme as depicted in Fig.~\ref{fig:scheme}.
Here, the relays recover degraded sets of the message parts in the natural order---recall the
assumption on the channel gains in~\eqref{eq:gains}---say, relay~$1$ recovers $(M_1,\ldots, M_N)$,
relay~$2$ recovers $(M_2, \ldots, M_N)$, relay~$3$ recovers $(M_3, \ldots, M_N)$, and so on. The relays then cooperatively communicate
these message parts to each destination
as in the multiple access channel with degraded message sets~\cite{Han1979, Prelov1984}.

\subsection{The Diamond Network}
For simplicity of exposition, we first consider the case $L = 1$.
The achievable rate of the proposed scheme can be characterized as
\begin{align*}
\Rmpdf =
\max \biggl\{ \sum_{j=1}^N R_j \suchthat (R_1, \ldots, R_N) \in \Rbcdms \cap \Rmacdms \biggr\},
\end{align*}
where $\Rbcdms$ is the capacity region of the standard $N$-receiver Gaussian broadcast channel (BC) with degraded message sets
and $\Rmacdms$ is the capacity region of the $N$-sender Gaussian multiple access channel (MAC)
with degraded message sets. Since the broadcast channel is degraded in the order of $1\to2 \to \cdots \to N$, $\Rbcdms = \Rbc$ as in \eqref{eq:Rbc}. The capacity region of the multiple access channel with degraded message sets~\cite{Han1979, Prelov1984} consists of all rate tuples
$(R_1,\ldots, R_N)$ such that
\begin{equation} \label{eq:Rmacdms}
%\sum_{j=1}^k R_j \le \Phi(1,k) ,\quad
%  k \in [1::N],
\sum_{j=1}^k R_j \le  I(\Xt^k; \Yt_{1} | \Xt_{k+1}^N),\quad
  k \in [1::N],
\end{equation}
for some $F(\xt^N)$ such that $\E(\Xt_j^2) \le P$, $j \in [1::N]$.
Again by the maximum differential entropy lemma, there is
no loss of generality in setting $(\Xt_1,\ldots, \Xt_N)$ to be jointly Gaussian
in~\eqref{eq:Rmacdms}.

In order to obtain a lower bound on $\Rmpdf$,
we follow the same approach \cite{Chern--Ozgur2012, Chern--Ozgur2014} as reviewed in the previous section and
use the polymatroidal inner bound on $\Rbcdms$ in~\eqref{eq:bc-ineqs}.
As for $\Rmacdms$, we note that the region in \eqref{eq:Rmacdms} is a polymatroid
for a fixed $F(\xt^N)$; cf.\@ Example~\ref{polymatroid-2}.
Thus, by Edmonds's polymatroid intersection theorem with
\begin{align}
\phi(\Jc) &= I(\Xt^{\Jc_{\max}}; \Yt_{1} | \Xt_{\Jc_{\max}+1}^N), \label{eq:mac-dms-ineqs} \\
\psi(\Jc) &= \C\biggl( \sfrac{1}{N} \sum_{j \in \Jc} S_j \biggr), \notag
\end{align}
the achievable rate of the proposed scheme is lower bounded as
\begin{align}
\Rmpdf &\ge
\sup_{F} \min_{\Jc \subseteq [1::N]} \bigl [ \psi(\Jc^c)+ \phi(\Jc) \bigr ], \label{eq:mpdf1}
\end{align}
where the supremum is over all jointly Gaussian $\Xt^N$ satisfying $\E(\Xt_j^2) \le P$, $j \in [1::N]$.
Since for each $\Jc \subseteq [1::N]$ with $\Jc_{\max} = k$,
\begin{align*}
\psi(\Jc^c) + \phi(\Jc) &\ge \psi([1::k]^c) + \phi(\Jc) \\
&= \psi([k+1::N]) + \phi([1::k]),
\end{align*}
the minimum in~\eqref{eq:mpdf1} is attained by $\Jc = \emptyset$ or $\Jc = [1::k]$ for some $k$. Thus,
\begin{equation}
\Rmpdf \ge
\sup_{F} \min_{k \in [0::N]} \biggl [ \C\biggl( \sfrac{1}{N}\! \sum_{j = k+1}^N S_j \biggr)
+ I(\Xt^k; \Yt_{1} | \Xt_{k+1}^N) \biggr ]. \label{eq:mpdf2}
\end{equation}
In comparison, by restricting $\Jc$ to be of the form $[1::k]$ in \eqref{eq:cutset1},
the cutset upper bound can be relaxed as
\begin{align}
\Rcs
%&\le
%\sup_{F} \min_{k \in [0::N]}  \biggl [ \psi^{'}([k+1::N])
%+ \Phi(1,k) \biggr ].
&\le
\sup_{F} \min_{k \in [0::N]}  \biggl [ \C\biggl( \sum_{j = k+1}^N S_j \biggr)
+ I(\Xt^k; \Yt_{1} | \Xt_{k+1}^N) \biggr ].
\label{eq:cutset3}
\end{align}
By comparing \eqref{eq:mpdf2} and \eqref{eq:cutset3}, we establish the following.

\begin{proposition}
The gap between the achievable rate of the proposed partial decode--forward
scheme and the cutset bound is upper bounded as
\[
\Deltaprimepdf
 := \Rcs - \Rmpdf \le \half \log N,
\]
regardless of $S_j$ and $\St_{1k}$, $j, k \in [1::N]$.
\end{proposition}

\subsection{The General Two-Hop Network}
\label{GMPDF}
The advantage of the modified partial decode--forward coding scheme is fully realized when there are multiple destinations ($L \ge 2$), in which case the Chern--\Ozgur{} scheme has an unbounded gap from the capacity \cite[Sec.~VI]{Chern--Ozgur2012}.
Recall from Fig.~\ref{fig:scheme} that
in the proposed partial decode--forward scheme, the message parts are communication
over a cascade of a BC (with degraded message sets) and multiple MACs with degraded message sets.
The achievable rate can be thus characterized as
\begin{align*}
\Rmpdf
&=
\max \biggl\{ \sum_{j=1}^N R_j \suchthat (R_1, \ldots, R_N) \in \Rbcdms \cap \Rmackdms \biggr\},
\end{align*}
where $\Rmackdms$ is the set of rate tuples $(R_1,\ldots, R_N)$ such that
\begin{equation} \label{eq:Rmmacdms}
%\sum_{j=1}^k R_j \le \min_{d \in [1::L]} \Phi(d,k),\quad
%  k \in [1::N],
\sum_{j=1}^k R_j \le \min_{d \in [1::L]} I(\Xt^k; \Yt_{d} | \Xt_{k+1}^N),\quad
  k \in [1::N],
\end{equation}
for some jointly Gaussian $\Xt^N$ with $\E(\Xt_j^2) \le P$, $j \in [1::N]$,
which is identical to the capacity region of the $N$-sender $L$-state Gaussian compound MAC
with degraded message sets.

We can now proceed in the exactly same manner as in the single-destination case, except that
in place of \eqref{eq:mac-dms-ineqs} we have another polymatroid
\[
\phi(\Jc)
%&\le \min_{d \in [1::L]} \Phi(d,\Jc_{max}), \quad
%\Jc \subseteq [1::N]. \label{eq:mac-dms-ineqs2}
= \min_{d \in [1::L]} I(\Xt^{\Jc_{\max}}; \Yt_{d} | \Xt_{\Jc_{\max}+1}^N).
%\label{eq:mac-dms-ineqs2}
\]
Consequently, we can lower bound the achievable rate of the scheme as
\begin{align}
&\Rmpdf  \notag \\
%\ge
%\sup_F \min_{d \in [1::L]} \min_{k\in[0::N]} \bigl [ \psi([k+1::N]) + \Phi(d,k) \bigr ], \label{eq:mpdf-multi}
&\ge
\sup_F \min_{d \in [1::L]} \min_{k\in[0::N]} \biggl [ \C\biggl( \sfrac{1}{N}\! \sum_{j = k+1}^N S_j \biggr)
+ I(\Xt^k; \Yt_{d} | \Xt_{k+1}^N) \biggr ].
\label{eq:mpdf-multi}
\end{align}
In comparison,
\begin{align*}
&\Rcs \notag \\
%\le
%\sup_F \min_{d \in [1::L]} \min_{k\in[0::N]} \biggl [ \psi^{'}([k+1::N])
%+ \Phi(d,k) \biggr ].
&\le
\sup_F \min_{d \in [1::L]} \min_{k\in[0::N]} \biggl [ \C\biggl( \sum_{j = k+1}^N S_j \biggr)
+ I(\Xt^k; \Yt_{d} | \Xt_{k+1}^N) \biggr ].
%\label{eq:cutset-multi}
\end{align*}
This establishes the following.

\begin{theorem}
The gap between the achievable rate of the proposed partial decode--forward
scheme and the cutset bound is upper bounded as
\[
\Delta'_\mathrm{PDF} = \Rcs - \Rmpdf \le \half \log N,
\]
regardless of the SNRs $S_j$ and $\St_{dk}$, $j, k \in [1::N]$, $d \in [1::L]$, and 
the number of destinations $L = 1,2, \ldots.$
\end{theorem}

A few remarks are in order.
\begin{enumerate}
\item When
$\Rmackdms \subseteq \Rbcdms$ (which is the case, for example, if $|g_N|\geq \min_d \sum_{j=1}^N|\gt_{dj}|$),
the proposed coding scheme actually achieves the capacity
\[
C= \min_{d \in [1::L]} \C\biggl( \Bigl(\sum_{j=1}^{N} \sqrt{\St_{dj}}\Bigr)^2 \biggr).
\]
In this case, the coding scheme simplifies to a simple decode--forward scheme, whereby every relay recovers the message $M$
and coherently forwards it.

\item At the other extreme, when $\Rbcdms \subseteq \Rmackdms$ (which is the case, for example, if $|g_1|\le \min_d |\gt_{d1}|$),
the maximum achievable rate of the proposed coding scheme is
\[
\Rmpdf = \C( S_1 ).
\]
Note that this rate is achieved trivially by using only the best relay (relay~$1$) and keeping the other relays idle,
yet the gap from the capacity is no more than $(1/2) \log N$.
\end{enumerate}

The performance difference between the Chern--\Ozgur{} PDF scheme and the proposed PDF scheme is best illustrated
by the following example taken from \cite[Sec.~VI]{Chern--Ozgur2012}.

\begin{example}
\label{ex:gap}
Consider the Gaussian two-hop relay network with $2$ relays and $2$ destinations as depicted in 
Fig.~\ref{multicast-example}, where the coefficients indicate the corresponding 
channel gains.
\begin{figure}[h]
\centering
\psfrag{S}{}
\psfrag{D1}{}
\psfrag{D2}{}
\psfrag{a}{$a$}
\psfrag{b}{$\sqrt{a}$}
\psfrag{c}{$0$}
\psfrag{R1}{}
\psfrag{R2}{}
\psfrag{E1}{$\C(a)$}
\psfrag{E2}{$\C(a^2)$}
\psfrag{E3}{$\C(a^2/2)$}
\psfrag{E4}{$\C(a/2)$}
\psfrag{E5}{$\C(a^2+a)$}
\psfrag{RM1}{$\Rr_\mathrm{MAC1}$}
\psfrag{RM2}{$\Rr_\mathrm{MAC2}$}
\psfrag{RB}{$\Rbc$}
\psfrag{RPDF}{$\Rpdf$}
\psfrag{RI1}{$\underline{\Rr}_\mathrm{MAC1}$}
\psfrag{RI2}{$\underline{\Rr}_\mathrm{MAC2}$}
\psfrag{RBI}{$\Rbcrelaxed$}
\psfrag{RMPDF}{$\Rmpdfrelaxed$}
\includegraphics[scale=0.7]{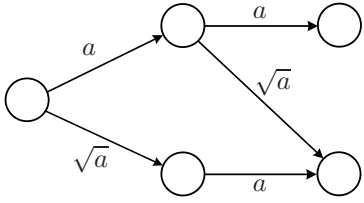}
%\subfigure[]{\includegraphics[scale=0.7]{figs/Example-result1-1.eps}}
%\subfigure[]{\includegraphics[scale=0.7]{figs/Example-result1-2.eps}}
\caption{An example network.}
\label{multicast-example}
\end{figure}
The cutset bound is bounded as 
\[
\C(a^2 P) \le \Rcs \le \C((a + \sqrt{a})^2 P),
\]
where the lower bound follows by setting $X, \Xt_1, \Xt_2$ to be independent $\N(0,P)$ in~\eqref{eq:cutset}
and the upper bound follows by considering only the broadcast cut.
The achievable rate of 
the PDF scheme by Chern and \Ozgur{} is 
\[
\Rpdf = \C(a P),
\]
which has an arbitrarily large gap from the cutset bound as $a \to \infty$.
In comparison, the achievable rate
of the proposed PDF scheme is lower bounded as
\[
\Rmpdf \ge \C\left(\frac{(a+a^2)P}{2}\right),
\]
which is within 1 bit from the cutset bound.
%The rates achieved by the Chern--\Ozgur{} scheme and our proposed scheme are sketched in Fig.~\ref{multicast-example}(b) 
%and Fig.~\ref{multicast-example}(c), respectively. 
%The capacity of this network is approximately $\C(a^2)$ when $a$ is large. To achieve the multicasting capacity within $1$ bit with our MPDF scheme would require the source node to split its message into two independent part $M_{1}$ and $M_{2}$ each of rate $R_{1}=C(a)$ and $R_{2}=C((a+a^2)/2)-C(a)$. The relay node $1$ decode both message, re-encode, and forward them to the destinations, while the relay node $2$ only decode $M_{2}$, re-encode, and forward it to the destinations. The rate achieved by using this scheme is $C((a+a^2)/2)$, which is only $1$ bit away from the cutset upper bound. On the other hand, 
%This is because even though relay node $1$ can decode both messages from the source, it only forward its desired message. In order to decode both messages at the destinations, we need to set $M_{2}=\emptyset$ and $M_{1}=M$. When $a$ is large, the PDF scheme proposed by Chern and \"Ozg\"ur will fail to achieve the multicasting capacity within the same gap.
\end{example}

\section{Linear-Complexity Capacity Approximation}
\label{Linear-complexity-app}

Computation of the achievable rate in~\eqref{eq:mpdf-multi} requires
maximization over all Gaussian input distributions $F$.
We now restrict the distribution to be
independent and identically distributed $\Xt_j \sim \N(0,P)$, $j \in [1::N]$.
This can be interpreted as a more practical coding scheme in which
the relays use independent Gaussian codebooks and transmit codewords noncoherently.
The achievable rate of the scheme is lower bounded by
\begin{align}
\Rmpdfsub
&\ge
\min_{d \in [1::L]} \min_{k\in[0::N]} \biggl [ \C\biggl( \sfrac{1}{N}\! \sum_{j = k+1}^N S_j \biggr)
+ \C\biggl( \sum_{j=1}^k \St_{dj} \biggr) \biggr ].
\label{eq:mpdf-multi-relaxed}
\end{align}
In comparison, starting with \eqref{eq:cutset2} and 
following the same argument as before,
we can relax the cutset upper bound as
\begin{align}
\Rcs
%&\le
%\min_{d \in [1::L]} \min_{k\in[0::N]} \biggl [ \psi^{'}([k+1::N])
%+ \C\biggl( \Bigl(\sum_{j=1}^k \sqrt{\St_{dj}}\Bigr)^2 \biggr) \biggr ] \notag \\
%&\le
%\min_{d \in [1::L]} \min_{k\in[0::N]} \biggl [ \psi^{'}([k+1::N])
%+ \C\biggl( N \sum_{j=1}^k \St_{dj} \biggr) \biggr ].
&\le
\min_{d \in [1::L]} \min_{k\in[0::N]} \biggl [ \C\biggl( \sum_{j=k+1}^N S_j \biggr)
+ \C\biggl( \Bigl(\sum_{j=1}^k \sqrt{\St_{dj}}\Bigr)^2 \biggr) \biggr ].
\label{eq:cutset-multi-relaxed}
\end{align}
Thus, by~\eqref{eq:gap-term1} and~\eqref{eq:gap-term2}, the capacity is approximated uniformly by $\log N$. 
Moreover, the computation of~\eqref{eq:mpdf-multi-relaxed}
or~\eqref{eq:cutset-multi-relaxed} involves computing Gaussian capacity functions for $L(N+1)$ cuts, which is
a significant savings from the directed computation of the cutset bound
with all $L2^N$ possible cuts as in \eqref{eq:cutset}.

We summarize this result as follows.
\begin{proposition}
The capacity of the Gaussian two-hop network is bounded as
\begin{align*}
C &\ge \min_{d \in [1::L]} \min_{k\in[0::N]} \biggl [ \C\biggl( \sfrac{1}{N}\! \sum_{j = k+1}^N S_j \biggr)
+ \C\biggl( \sum_{j=1}^k \St_{dj} \biggr) \biggr ], \\
C &\le
\min_{d \in [1::L]} \min_{k\in[0::N]} \biggl [ \C\biggl( \sum_{j=k+1}^N S_j \biggr)
+ \C\biggl( \Bigl(\sum_{j=1}^k \sqrt{\St_{dj}}\Bigr)^2 \biggr) \biggr ],
\end{align*}
where the gap between the lower and upper bounds is no greater than $\log N$
for any $S_j$ and $\St_{dk}$, $j, k \in [1::N]$, $d \in [1::L]$, and any $L$. Moreover,
both bounds can be computed in $O(LN)$ complexity.
\end{proposition}

These bounds yields a simple approximate expression for the capacity.
\begin{proposition}
\[
C = \min_{d \in [1::L]} \min_{k\in[0::N]} \biggl [ \C\biggl( \sum_{j = k+1}^N S_j \biggr)
+ \C\biggl( \sum_{j=1}^k \St_{dj} \biggr) \biggr ] \pm \half\log N.
\]
\end{proposition}

\section{Distributed Decode--Forward}
\label{DDF}
In this section, we consider the distributed decode--forward (DDF) coding scheme in~\cite{Lim--Kim--Kim--1--2014},
which is an extension of partial decode--forward to general multicast networks.
In particular, the rate achieved by DDF for our two-hop network is characterized \cite{Lim--Kim--Kim--1--2014} as 
\begin{align}
\Rddf =
\sup_{F} \min_{d,\Jc} &\Big[I(X, \Xt(\Jc) ; U(\Jc^c), \Yt_{d} | \Xt(\Jc^c)) \notag \\
&\quad - \sum_{k\in \Jc^c} I(U_k; X, \Xt^N | Y_k)\Big],
\label{eq:Rddf-multi}\end{align}
where the supremum is over all distributions of the form $(\prod_{k=1}^NF(\xt_k))F(x | \xt^N)F(u^N | x, \xt^N)$ satisfying $\E(X^2) \le P$ and $\E(\Xt_j^2) \le P$, $j \in [1::N]$.
By setting $X$ and $\Xt_j$ to be i.i.d. $\N(0,P)$ and
\begin{equation}
  U_j = Y_j - Z_j + \Zh_j, \quad j \in [1::N],
  \label{eq:choice-of-u}
\end{equation}
where $\Zh_j\sim \N(0,1)$, $j \in [1::N]$, are independent of each other and of $(\Xt^N,Y^N)$,
it can be shown \cite{Lim--Kim--Kim--1--2014} that
the gap between the achievable rate in \eqref{eq:Rddf-multi} and the cutset bound in \eqref{eq:cutset2} 
is no greater than $N/2$.

We now exploit the layered structure of the network to improve this $O(N)$ gap to $O(\log N)$.
Following a similar (and in some sense dual) development for noisy network coding in \cite{Chern--Ozgur2012},
we set $\Zh_j\sim \N(0,N)$ in \eqref{eq:choice-of-u}. Then, the first term of \eqref{eq:Rddf-multi} becomes
\begin{align*}
  &I(X, \Xt(\Jc); U(\Jc^c), \Yt_{d} | \Xt(\Jc^c))\\
  &\qquad\stackrel{(a)}{=} I(X; U(\Jc^c)) + I(\Xt(\Jc); \Yt_{d} | \Xt(\Jc^c))\\
  &\qquad= \C\biggl( \frac{1}{N} \sum_{j \in \Jc^c} S_j \biggl) +\C\biggl( \sum_{j \in \Jc} \St_{dj} \biggl),
\end{align*}
where $(a)$ follows by the independence of $(X, U^N)$ and $\Xt^N$ and the layered structure of the network.
For $k\in [1::N]$, each summand in the second term of \eqref{eq:Rddf-multi} becomes
\begin{align*}
  I(U_k; X, \Xt^N | Y_k) &= \half \log \biggl(\frac{1+\big(1+\frac{1}{N}\big)S_k}{1+S_k}\biggl)\\
   &\le \half \log \bigg(1+\frac{1}{N}\bigg)\\
   &\le \frac{1}{2N}.
\end{align*}
Hence, 
\begin{align}
\label{eq:Opt-Rddf-multi}
  & \Rddf \ge \min_{d,\Jc} \biggl [ \C\biggl( \frac{1}{N} \sum_{j \in \Jc^c} S_j \biggl) +\C\biggl( \sum_{j \in \Jc} \St_{dj} \biggl)- \frac{|\Jc^c|}{2N} \biggr].
\end{align}
Comparing this achievable rate in \eqref{eq:Opt-Rddf-multi} and the cutset bound in \eqref{eq:cutset2} 
establishes the following.

\begin{proposition}
\label{prop:ddf}
The gap between the achievable rate of the distributed decode--forward
scheme and the cutset bound is upper bounded as
\[
\Delta_\mathrm{DDF}=\Rcs - \Rddf \le \log N + \half,
\]
regardless of the SNRs $S_j$ and $\St_{dk}$, $j, k \in [1::N]$, $d \in [1::L]$, and 
the number of destinations $L = 1,2, \ldots.$
\end{proposition}

\section{Concluding Remarks}
\label{conclusion}

Multiple coding schemes achieve the multicast capacity of the two-hop Gaussian network with one source, $N$ relays,
and $L$ destinations within $O(\log N)$, including:
\begin{enumerate}
\item Noisy network coding (see~\cite[Th.~3.1]{Chern--Ozgur2014})

\item Distributed decode--forward (Prop.~\ref{prop:ddf} in the current paper)

\item Partial decode--forward (see~\cite[Th.~3.3]{Chern--Ozgur2014} for $L = 1$)

\item Partial decode--forward with degraded message sets (Th.~1 in the current paper).
\end{enumerate}
Among these, the fourth scheme, which is the main contribution of the paper,
achieves the tightest gap of $(1/2)\log N$ from the cutset bound. Moreover,
a simple lower bound on its achievable rate can be expressed as the minimum
of $L(N+1)$ cut rates, providing a sharp approximation of the capacity that
can be computed in $O(LN)$ complexity.
While it remains to be seen whether this linear-complexity approximation can be alternatively 
established via algebraic or combinatorial techniques, it is refreshing to note that
the best \emph{computational} result is obtained by a purely information-theoretic argument,
based directly on a simple coding scheme.

%\bibliographystyle{IEEEtran}
%\bibliography{nit}

\begin{thebibliography}{10}
\providecommand{\url}[1]{#1}
\csname url@samestyle\endcsname
\providecommand{\newblock}{\relax}
\providecommand{\bibinfo}[2]{#2}
\providecommand{\BIBentrySTDinterwordspacing}{\spaceskip=0pt\relax}
\providecommand{\BIBentryALTinterwordstretchfactor}{4}
\providecommand{\BIBentryALTinterwordspacing}{\spaceskip=\fontdimen2\font plus
\BIBentryALTinterwordstretchfactor\fontdimen3\font minus
  \fontdimen4\font\relax}
\providecommand{\BIBforeignlanguage}[2]{{%
\expandafter\ifx\csname l@#1\endcsname\relax
\typeout{** WARNING: IEEEtran.bst: No hyphenation pattern has been}%
\typeout{** loaded for the language `#1'. Using the pattern for}%
\typeout{** the default language instead.}%
\else
\language=\csname l@#1\endcsname
\fi
#2}}
\providecommand{\BIBdecl}{\relax}
\BIBdecl

\bibitem{Schein--Gallager2000}
B.~Schein and R.~G. Gallager, ``The {G}aussian parallel relay channel,'' in
  \emph{Proc. {IEEE} Int. Symp. Inf. Theory}, Sorrento, Italy, Jun. 2000,
  p.~22.

\bibitem{Schein2001}
B.~Schein, ``Distributed coordonation in network information theory,'' {Ph.D.}
  Thesis, Massachusetts Institute of Technology, Cambridge, USA, Oct. 2001.

\bibitem{El-Gamal1981b}
A.~El~Gamal, ``On information flow in relay networks,'' in \emph{Proc. IEEE
  National Telecomm. Conf.}, New Orleans, LA, Nov. 1981, vol.~2, pp.
  D4.1.1--D4.1.4.

\bibitem{Cover--El-Gamal1979}
T.~M. Cover and A.~El~Gamal, ``Capacity theorems for the relay channel,''
  \emph{{IEEE} Trans. Inf. Theory}, vol.~25, no.~5, pp. 572--584, Sep. 1979.

\bibitem{Avestimehr--Diggavi--Tse2011}
A.~S. Avestimehr, S.~N. Diggavi, and D.~N.~C. Tse, ``Wireless network
  information flow: {A} deterministic approach,'' \emph{{IEEE} Trans. Inf.
  Theory}, vol.~57, no.~4, pp. 1872--1905, Apr. 2011.

\bibitem{Lim--Kim--El-Gamal--Chung2011}
S.~H. Lim, Y.-H. Kim, A.~El~Gamal, and S.-Y. Chung, ``Noisy network coding,''
  \emph{{IEEE} Trans. Inf. Theory}, vol.~57, no.~5, pp. 3132--3152, May 2011.

\bibitem{Yassaee--Aref2011}
M.~H. Yassaee and M.~R. Aref, ``Slepian--{W}olf coding over cooperative relay
  networks,'' \emph{{IEEE} Trans. Inf. Theory}, vol.~57, no.~6, pp. 3462--3482,
  June 2011.

\bibitem{Chern--Ozgur2012}
B.~Chern and A.~\"{O}zg\"{u}r, ``Achieving the capacity of the $n$-relay
  {G}aussian diamond network within $\log n$ bits,'' in \emph{Proc. {IEEE} Inf.
  Theory Workshop}, Lausanne, Switzerland, Sep 2012, pp. 377--380.

\bibitem{Bergmans1974}
P.~P. Bergmans, ``A simple converse for broadcast channels with additive white
  {G}aussian noise,'' \emph{{IEEE} Trans. Inf. Theory}, vol.~20, no.~2, pp.
  279--280, 1974.

\bibitem{Prelov1984}
V.~V. Prelov, ``Transmission over a multiple-access channel with a special
  source hierarchy,'' \emph{Probl. Peredachi Inf.}, vol.~20, no.~4, pp. 3--10,
  1984.

\bibitem{Han1979}
T.~S. Han, ``The capacity region of general multiple-access channel with
  certain correlated sources,'' \emph{Inf. Control}, vol.~40, no.~1, pp.
  37--60, 1979.

\bibitem{Parvaresh--Etkin2011}
F.~Parvaresh and R.~H. Etkin, ``On computing the capacity of relay networks in
  polynomial time,'' in \emph{Proc. {IEEE} Int. Symp. Inf. Theory}, Palo Alto,
  Carlifornia, July/Aug. 2011, pp. 1342--1346.

\bibitem{Brahma--Sengupta--Fragouli2014}
S.~Brahma, A.~Sengupta, and C.~Fragouli, ``Efficient subnetwork selection in
  relay networks,'' in \emph{Proc. {IEEE} Int. Symp. Inf. Theory}, Honolulu,
  HI, June/July 2014, pp. 1927--1931.

\bibitem{Lim--Kim--Kim--1--2014}
S.~Lim, K.~T. Kim, and Y.-H. Kim, ``Distribued decode--forward for
  multicasting,'' in \emph{Proc. {IEEE} Int. Symp. Inf. Theory}, Honolulu, HI,
  June/July 2014, pp. 636--640.

\bibitem{Lim--Kim--Kim--2--2014}
------, ``Distribued decode--forward for broadcast,'' in \emph{Proc. {IEEE}
  Inf. Theory Workshop}, Hobart, TAS, November 2014, pp. 556--560.

\bibitem{El-Gamal--Kim2011}
A.~El~Gamal and Y.-H. Kim, \emph{Network Information Theory}.\hskip 1em plus
  0.5em minus 0.4em\relax Cambridge: Cambridge University Press, 2011.

\bibitem{Schrijver2003}
A.~Schrijver, \emph{Combinatorial Optimization. {\em 3 vols}}.\hskip 1em plus
  0.5em minus 0.4em\relax Berlin: Springer-Verlag, 2003.

\bibitem{Edmonds1970}
J.~Edmonds, ``Submodular functions, matroids, and certain polyhedra,'' in
  \emph{Combinatorial Structures and Their Applications}, Gordon and Breach,
  New York, 1970, pp. 69--87.

\bibitem{Chern--Ozgur2014}
B.~Chern and A.~\"Ozg\"ur, ``Achieving the capacity of the $n$-relay {G}aussian
  diamond network within $\log n$ bits,'' \emph{{IEEE} Trans. Inf. Theory},
  vol.~60, no.~12, pp. 7708--7718, Dec. 2014.

\bibitem{Vishwanath--Jindal--Goldsmith2003}
S.~Vishwanath, N.~Jindal, and A.~J. Goldsmith, ``Duality, achievable rates, and
  sum-rate capacity of {G}aussian {MIMO} broadcast channels,'' \emph{{IEEE}
  Trans. Inf. Theory}, vol.~49, no.~10, pp. 2658--2668, 2003.

\end{thebibliography}
\newcommand{\noopsort}[1]{}

\end{document}